\documentclass{jfm}

\usepackage{amsmath,amssymb}
\usepackage{graphicx,tikz,pgfplots}
\graphicspath{{figs/},{frames/}}
\usepackage{natbib}

\newcommand{\lan}{\left\langle}
\newcommand{\ran}{\right\rangle}
\newcommand{\bu}{\mathbf{u}}

\newcommand{\wT}{\lan wT \ran}
\newcommand{\T}{\lan T \ran}

\newcommand{\oT}{\overline{T}}

\newcommand{\Tm}{\oT_{max}}

\newcommand{\fB}{\mathcal F_B}
\newcommand{\wt}{\widetilde}
\newcommand{\e}{\times10^}

\newcommand\solidrule[1][21pt]{\rule[0.5ex]{#1}{1pt}}
\newcommand\dashedrule{\mbox{%
	\solidrule[5pt]\hspace{3pt}\solidrule[5pt]\hspace{3pt}\solidrule[5pt]}}

\begin{document}

\shorttitle{Penetrative internally heated convection}
\shortauthor{D.\ Goluskin and E.\ P.\ van der Poel}

\title{Penetrative internally heated convection\\in two and three dimensions}

\author
 {
 David Goluskin\aff{1}
  \corresp{\email{goluskin@umich.edu}} \and
  Erwin P. van der Poel\aff{2}
  }

\affiliation
{
\aff{1}
Mathematics Department and Center for the Study of Complex Systems,\\University of Michigan, Ann Arbor, MI, USA
\aff{2}
Physics of Fluids Group, Faculty of Science and Technology, J.M. Burgers Center for Fluid Dynamics and MESA+ Institute, University of Twente, Enschede, The Netherlands
}

\maketitle

\begin{abstract}
\noindent Convection of an internally heated fluid, confined between top and bottom plates of equal temperature, is studied by direct numerical simulation in two and three dimensions. The unstably stratified upper region drives convection that penetrates into the stably stratified lower region. The fraction of produced heat escaping across the bottom plate, which is one half without convection, initially decreases as convection strengthens. Entering the turbulent regime, this decrease reverses in two dimensions but continues monotonically in three dimensions. The mean fluid temperature, which grows proportionally to the heating rate ($H$) without convection, grows proportionally to $H^{4/5}$ when convection is strong in both two and three dimensions. The ratio of the heating rate to the fluid temperature is likened to the Nusselt number of Rayleigh-B\'enard convection. Simulations are reported for Prandtl numbers between 0.1 and 10 and for Rayleigh numbers (defined in terms of the heating rate) up to $5\times10^{10}$.
\end{abstract}

\section{Introduction}

Internally heated (IH) convection refers to fluid motion driven by buoyancy forces that arise when a fluid is heated by sources within its volume. This differs from the more-studied phenomenon of Rayleigh-B\'enard (RB) convection, which is driven by thermal conditions at the fluid boundaries. IH convection occurs in many systems, driven by various heating mechanisms. Astrophysical and geophysical mechanisms include nuclear fusion in the cores of large stars \citep{Kippenhahn1994}, radioactive decay in the Earth's mantle \citep{Schubert2001}, Kelvin-Helmholtz heating due to gravitational contraction in gas giants and brown dwarfs \citep{Irwin2009}, and tidal heating of planets and moons \citep{Peale1979}. In engineered systems, IH convection can occur when chemical or nuclear reactions occur in a fluid, or when viscous dissipation heats a turbulent flow. Although we speak only of heating, similar dynamics govern convection driven by internal cooling (as in a radiating atmosphere) or by sources or sinks of other scalars (as created by reactions).

\begin{figure}
\begin{center}
(a)
\begin{tikzpicture}
\node at (0,0) {\includegraphics[width=103pt]{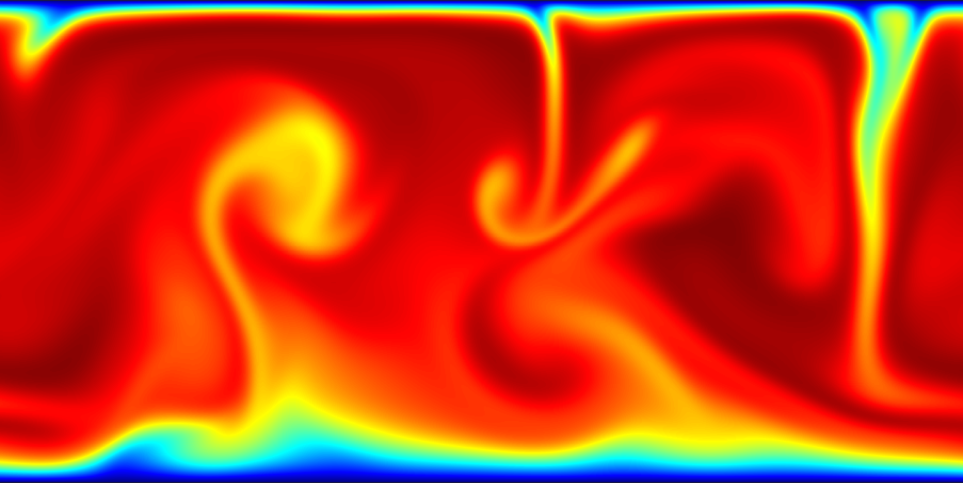}};
\foreach \x in {-1.8,-1.6,...,-.8} \draw[thick,black] (\x,-.94) -- (\x+.2,-1.24);
\foreach \x in {.6,.8,...,1.6} \draw[thick,black] (\x,-.94) -- (\x+.2,-1.24);
\foreach \x in {-1.8,-1.6,...,-.8} \draw[thick,black] (\x,1.24) -- (\x+.2,.94);
\foreach \x in {.6,.8,...,1.6} \draw[thick,black] (\x,1.24) -- (\x+.2,.94);
\draw[line width=2pt] (-1.81,.94) -- (1.81,.94);
\draw[line width=2pt] (-1.81,-.94) -- (1.81,-.94);
\node at (0,1.14) {$T=0$};
\node at (0,-1.14) {$T=0$};
\draw[thick,-latex] (2.1,.9) -- (2.1,.3);
\node at (2.3,.7) {$\mathbf{g}$};
\end{tikzpicture}
(b)
\includegraphics[width=100pt]{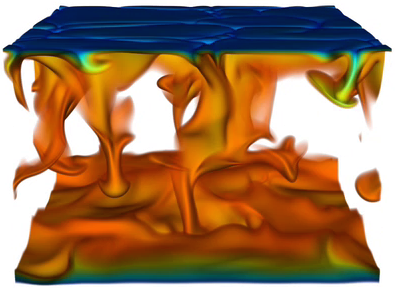}
\end{center}
\caption{\label{fig:config} (a) Schematic of the present convective model, which is subject to gravitational acceleration $\mathbf{g}$ and uniform internal heating. Both boundary plates are fixed at a temperature defined as zero. Example temperature fields are shown for simulations in 2D and (b) in 3D. The colour scale indicates fluid that is cooler at the boundaries and warmer in the interior, and in panel \emph{b} the hottest fluid is transparent to aid visualization. Control parameters for both simulations are $Pr=1$ and $R=5\times10^8$. A movie of the 3D simulation is provided online.}
\end{figure}
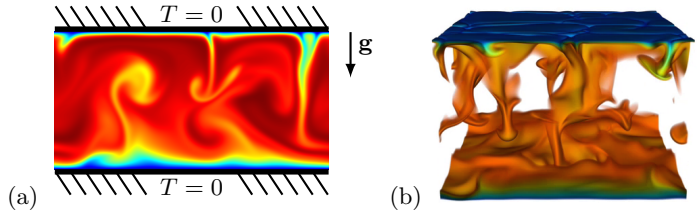

Figure \ref{fig:config}a shows the IH configuration we consider here: a horizontal layer of fluid, bounded above and below by plates of fixed and equal temperatures, that is subject to constant and uniform heating throughout its volume. These boundary conditions are especially relevant to systems that are cooled above and below, such as liquid metal batteries \citep{Shen2015} or overreactions in nuclear reactor accidents \citep{Asfia1996, Nourgaliev1997, Grotzbach1999}. IH convection in this configuration and others is reviewed by \citet{Goluskin2015a}. 

In the present configuration, all the fluid is hotter than the bounding plates, so heat flows outward across both boundaries. The warmer and more buoyant fluid in the centre of the layer is sandwiched between cooler and less buoyant fluid near the boundaries, so density is unstably stratified near the top and stably stratified near the bottom---a situation analogous to counter-rotating Taylor-Couette flow \citep{Ostilla-Monico2014} and rotating pipe flow \citep{Orlandi1997}. This results in penetrative convection, meaning that motions driven by buoyancy forces in the unstable upper region penetrate into the stable lower region. Such convective motions are illustrated by figure \ref{fig:config}, which shows temperature fields from two of the simulations described below.

RB convection is the canonical simplification of convection driven by thermal boundary conditions. The present configuration, which is no less fundamental, can likewise be regarded as the canonical simplification of penetrative convection driven by internal sources or sinks. The primary questions about this configuration that motivate the present study are: how hot is the fluid, and in what proportions does the internally produced heat escape across the top and bottom boundaries, respectively? Answers to such questions depend on the dimensionless control parameters: the usual Prandtl number ($Pr$) and a Rayleigh number ($R$) defined below in terms of the heating rate.

Quantitative studies of the present configuration have included laboratory experiments in which the fluid was heated by electric current \citep{Kulacki1972, Jahn1974, Ralph1977} or fixed heating elements \citep{Lee2007}, as well as direct numerical simulations (DNS) in 2D \citep{Emara1980, Jahn1974, Goluskin2012} and in 3D \citep{Grotzbach1988, Worner1997}. Most of these studies have varied $R$ with $Pr\approx7$, and they have paid particular attention to the maximum of the horizontally or temporally averaged temperature. The ratio of this maximum temperature to the heating rate has been found to decrease at rates between $R^{-0.18}$ and $R^{-0.22}$ as $R$ is raised \citep{Goluskin2015a}. Mean fluid temperature, rather than a maximum temperature, was studied in the 2D DNS of \citet{Goluskin2012}, where $R$ was varied at several fixed $Pr$. The ratio of this mean temperature to the heating rate was found to decrease proportionally to $R^{-0.20}$. Additionally, the fraction of heat escaping across the bottom boundary, as opposed to the top one, was found to initially fall but then increase as $R$ was raised. Such non-monotonicity has not been reported in 3D.

In the present work, we extend the 2D DNS data of \citet{Goluskin2012} by sweeping though $Pr$ at fixed $R$, and we carry out 3D DNS over a similar parameter range, conducting one sweep though $R$ and one through $Pr$. We report on the mean fluid temperature, a mean maximum temperature, and the asymmetry between heat fluxes across the top and bottom boundaries.

Section \ref{sec: eq} defines the governing model. Section \ref{sec: int quant} then discusses the integral quantities most important to heat transport, including past findings and key questions. Section \ref{sec: DNS} presents DNS results, along with an analogy between the inverse mean fluid temperature and the Nusselt number of RB convection. Concluding remarks appear in \S\ref{sec: con}, and data are tabulated in the Appendix.

\section{Governing equations}
\label{sec: eq}

For the governing model we adopt the Oberbeck-Boussinesq approximation, in which the fluid has constant kinematic viscosity, $\nu$, thermal diffusivity, $\kappa$, and coefficient of thermal expansion, $\alpha$. We nondimensionalize length by the layer height, $d$, time by the thermal diffusion timescale, $d^2/\kappa$, and temperature by $d^2H/\kappa$, where $H$ is the heating rate in units of temperature per time. The dimensionless Boussinesq equations governing the velocity $\mathbf{u}=(u,v,w)$, temperature $T$, and pressure $p$ are then \citep{Rayleigh1916, Chandrasekhar1981}
\begin{align}
\nabla \cdot \mathbf u &= 0, \label{eq: inc} \\
\partial_t \mathbf u + \mathbf u \cdot \nabla \mathbf u &=
	-\nabla p + Pr \nabla^2 \mathbf u + PrR\,T\,\hat{\mathbf z}, \label{eq: u} \\
\partial_t T + \mathbf u \cdot \nabla T &= \nabla^2 T + 1. \label{eq: T}
\end{align}
The dimensionless control parameters are the Rayleigh and Prandtl numbers defined by
\begin{align}
R &= \frac{g\alpha d^5 H}{\kappa^2\nu}, & Pr &= \frac{\nu}{\kappa}, \label{eq: R}
\end{align}
where $g$ is the uniform gravitational acceleration in the $-\hat{\mathbf z}$ direction. The parameter $R$ is standard in the study of IH convection but differs from the Rayleigh number of RB convection, where the temperature scale comes from the boundary conditions instead of the heating rate.

The dimensionless vertical extent is $-1/2\le z\le 1/2$. At the boundaries we impose no-slip conditions on the velocity, and we fix the temperatures to zero (without loss of generality), so $\bu,T = 0$ at $z=\pm1/2$. Both horizontal directions are periodic and have the same aspect ratio, $\Gamma$, so the horizontal coordinates are bounded by $0\le x,y<\Gamma$. As described below, we choose $\Gamma$ large enough so that the global quantities of interest are not sensitive to $\Gamma$.

\section{Integral quantities}
\label{sec: int quant}

We are especially interested in the mean fluid temperature and the fractions of internally produced heat escaping across the top and bottom boundaries. These quantities and additional information about vertical structure are captured by the mean temperature profile, $\oT(z)$, where an overbar denotes an average over the horizontal directions and infinite time. (Such infinite-time averages are approximated in simulations by long finite times.) When the fluid is static, the equilibrium temperature profile is parabolic: $T=\tfrac{1}{8}(1-4z^2)$. When $R$ is too small to sustain convection, all initial conditions evolve toward this static state, which is parameter-independent by virtue of the nondimensionalization. When there is no constraint on horizontal wavenumbers, convection is guaranteed by linear instability of the static state when $R>37\,325$ \citep{Sparrow1964}, and subcritical convection is possible at smaller $R$ \citep{Tveitereid1977}. Here we simulate large $R$, well above the onset of motion.

\begin{figure}
\centering
\includegraphics[height=120pt,trim=0 5 0 10,]{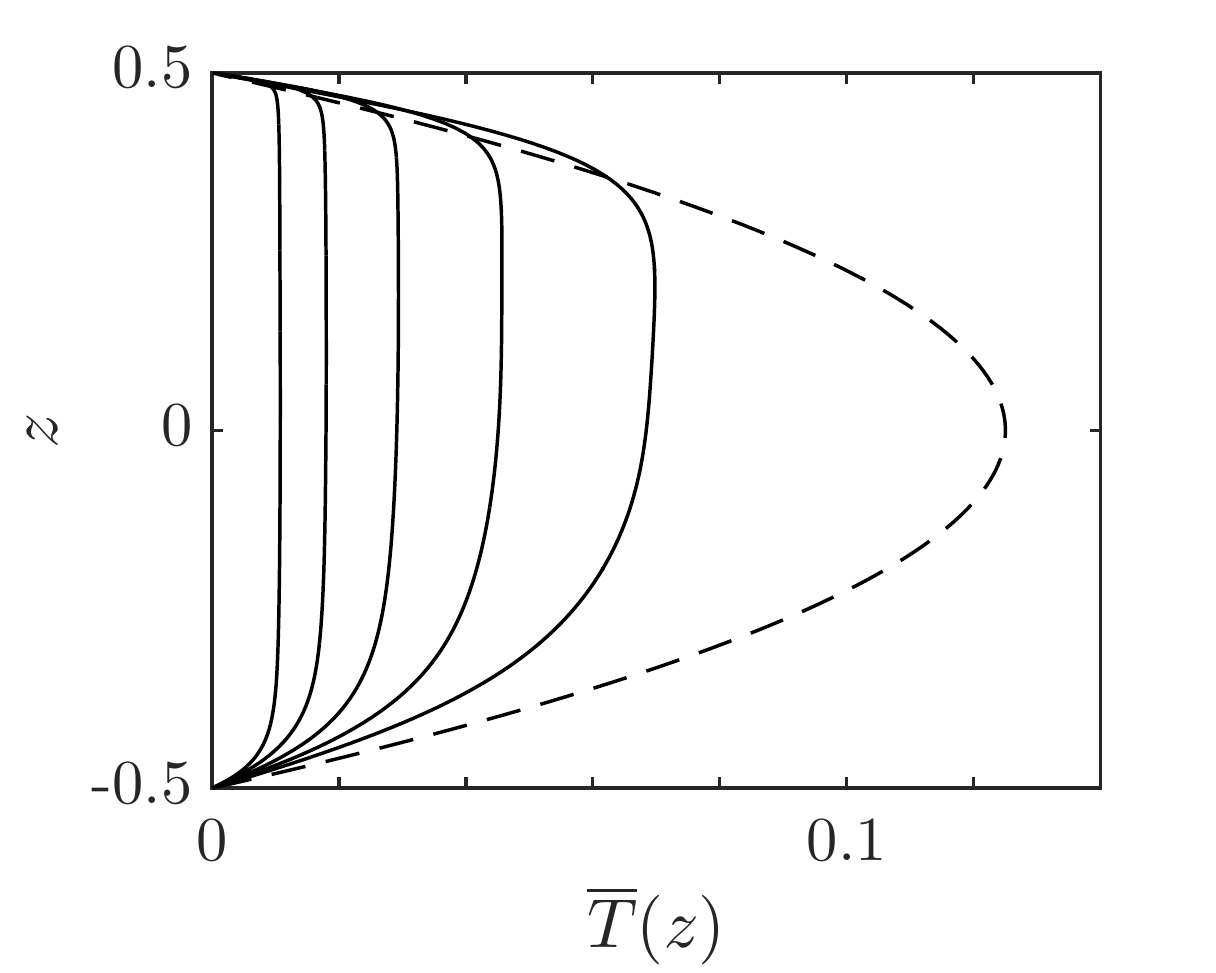}
\caption{\label{fig2} Mean vertical temperature profiles, $\oT(z)$, in the static state (\dashedrule) and in 3D simulations with $Pr=1$ and $R=10^6,10^7,10^8,10^9,10^{10}$ (from right to left).}
\end{figure} 

Figure \ref{fig2} shows the parabolic temperature profile of the static state---that is, the purely conductive state---along with selected mean temperature profiles, $\oT(z)$, for the 3D simulations described below. The basic qualitative trends in figure \ref{fig2} are shared by all temperature profiles reported for past simulations \citep{Peckover1974, Mayinger1975, Straus1976, Emara1980, Grotzbach1988, Worner1997, Goluskin2012} and experiments \citep{Kulacki1972, Mayinger1975, Ralph1977, Lee2007}. As $R$ is raised, convection further assists conduction in carrying heat to the boundaries. This decreases the dimensionless fluid temperature and brings the interior closer to isothermal. Thermal boundary layers form at the top and bottom, but the unstable top layer is thinner than the stable bottom layer, reflecting the fact that the majority of internally produced heat escapes across the top boundary.

To quantify the reduction of temperature by convection, we consider the (dimensionless) mean fluid temperature, $\T$, where angle brackets denote an average over volume and infinite time. Most researchers have instead considered $\Tm$, the maximum value of $\oT(z)$. The two quantities behave similarly, but $\T$ is more amenable to mathematical analysis, whereas $\Tm$ is easier to measure in the laboratory.

The mean temperature for any solution to the model (\ref{eq: inc})-(\ref{eq: T}) has been proven to obey the bounds \citep{Lu2004}
\begin{equation}
1.08\,R^{-1/3}<\T\le\tfrac{1}{12} \label{eq: T bounds}
\end{equation}
at large $R$. The quantity $\T$ saturates its upper bound of $1/12$ only in the static state and typically decreases as $R$ is raised. This decrease can be no faster than $\mathcal O(R^{-1/3})$ as $R\to\infty$. Since $T$ has been nondimensionalized using the heating rate, $H$, this means that the \emph{dimensional} mean temperature must grow with $H$ no slower than $\mathcal O(H^{2/3})$. In the only data reported for $\T$, which are from 2D DNS, $\T$ decays proportionally to $R^{-1/5}$ \citep{Goluskin2012}. Other studies report similar $R$-dependence for $\Tm$ \citep{Goluskin2015a}. Questions motivating our present study include: how different are $\T$ and $\Tm$, will a novel scaling of $\T$ be found at larger $R$, and how does $\T$ depend on $Pr$?

To quantify the up-down asymmetry that convection induces we examine $\fB$, the fraction of internally produced heat that flows outward across the bottom boundary. This fraction is related \emph{a priori} to the mean convective transport, $\wT$, by $\fB=1/2-\wT$ \citep{Goluskin2012}. Likewise, the fraction of heat flowing outward across the \emph{top} boundary must equal $1/2+\wT$ since the fractions crossing the top and bottom boundaries sum to unity. The quantity $\fB$ is not well understood, seemingly having no analogue in convection that is not penetrative, and only the crude bounds $0<\fB\le\tfrac{1}{2}$ have been proven mathematically \citep{Goluskin2012}.

Heat flows outward across the top and bottom boundaries in equal proportion in the static state, meaning that $\fB=1/2$. In sustained convection $\fB<1/2$ because the mean work exerted by buoyancy forces, which must be positive, is proportional to $\wT=1/2-\fB$. In all past studies, $\fB$ slowly decreases after the onset of convection as $R$ is raised over several decades \citep{Goluskin2015a}. At larger $R$, there is stark disagreement between past studies. In 2D simulations with $Pr=1$, $\fB$ reaches a minimum value of 0.33 and then \emph{increases} as $R$ is raised further \citep{Goluskin2012}. In experiments \citep{Kulacki1972} and 3D simulations \citep{Worner1997}, on the other hand, $\fB$ continues to decrease at the largest $R$ studied. Questions raised by these findings include: are the differing observations of $\fB$ explained by the differences between 2D and 3D flows, and what is the ultimate limit of $\fB$ as $R\to\infty$?

\section{Simulation results}
\label{sec: DNS}

We have simulated the governing equations (\ref{eq: inc})-(\ref{eq: T}) using an energy-conserving finite difference method \citep{Verzicco1996, Verzicco2003, VanderPoel2015}. Time averages are deemed converged when values of $\T$ and $\wT$ over the full post-transient duration agree with their values over half that duration to within 1\%. One way we have checked spatial resolution is from the integral balances $R\langle wT \rangle = \langle | \nabla {\bf u} |^2 \rangle$ and $\langle T \rangle = \langle | \nabla T |^2 \rangle$, derived by integrating $\bu\cdot(\ref{eq: u})$ and $T\times(\ref{eq: T})$, respectively. The balances are satisfied to within 1\% for all simulations reported here. We have also checked resolution by repeating some simulations at lower resolution and repeating every 3D simulation with $Pr=1$ and $R\le2\e8$ using the spectral element code {\tt nek5000} \citep{nek}. In each case, the {\tt nek5000} values of $\T$ and $\wT$ agree with the finite difference values to within 1\%. The Appendix gives further details on convergence, including time spans, meshes, and the resolution of boundary layers and Kolmogorov length scales. We have chosen aspect ratios large enough such that raising $\Gamma$ by approximately 50\% changes $\T$ and $\langle wT \rangle$ by less than 1\%. As shown in the Appendix, the necessary $\Gamma$ values decrease as $R$ is raised, from $\Gamma=\pi$ when $R\le 2\times10^6 $ to $\Gamma=1$ when $R\ge 2\e9$.

\begin{figure}
\centering
\begin{tikzpicture}
\node at (0,0) {\includegraphics[width=185pt]{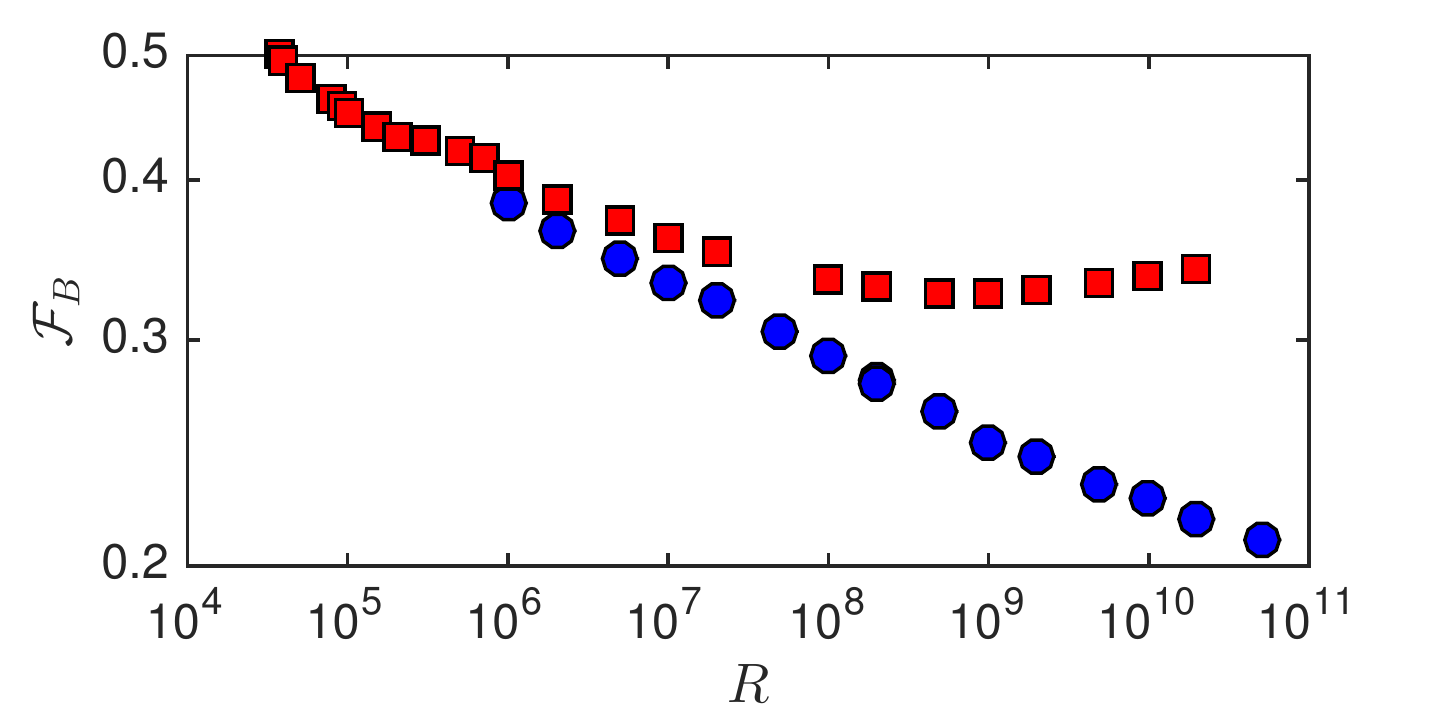}};
\node at (6.5,0) {\includegraphics[width=189pt]{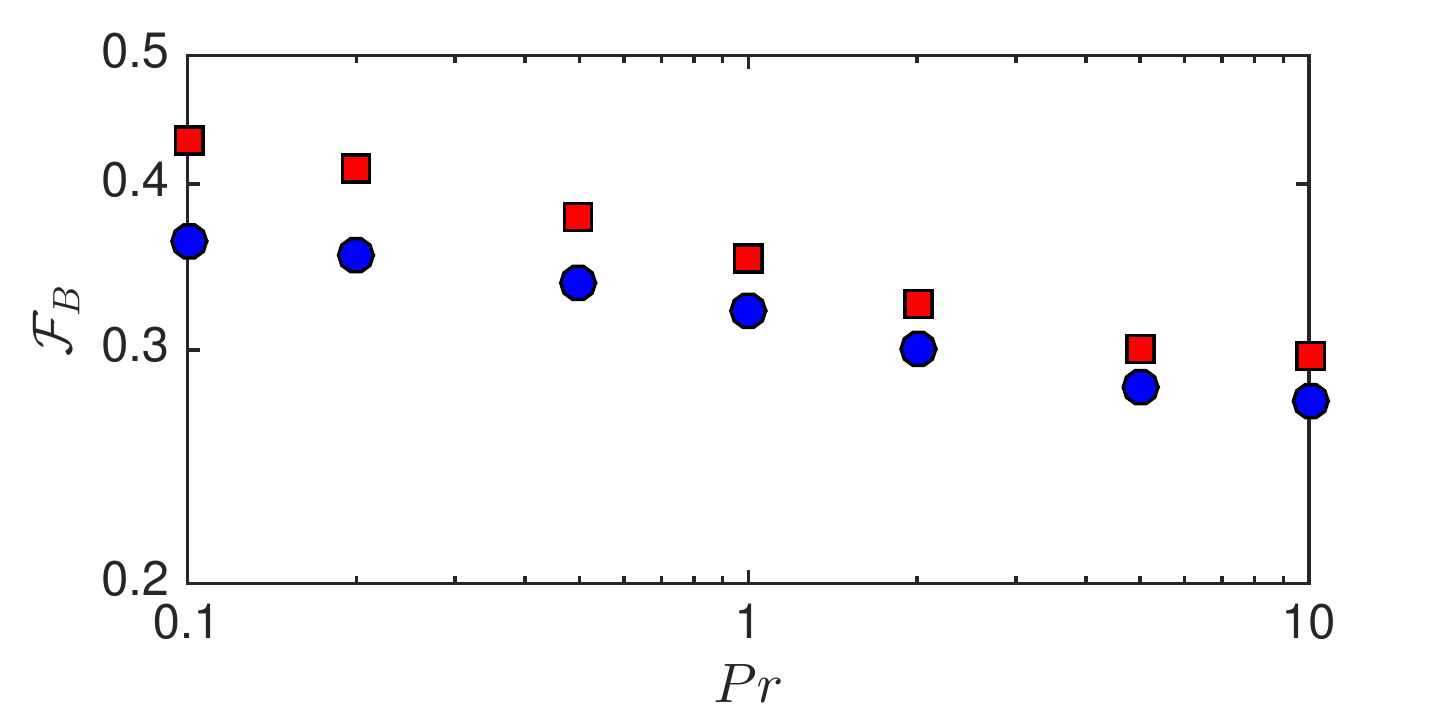}};
\node at (0,-2.955) {\includegraphics[width=185pt]{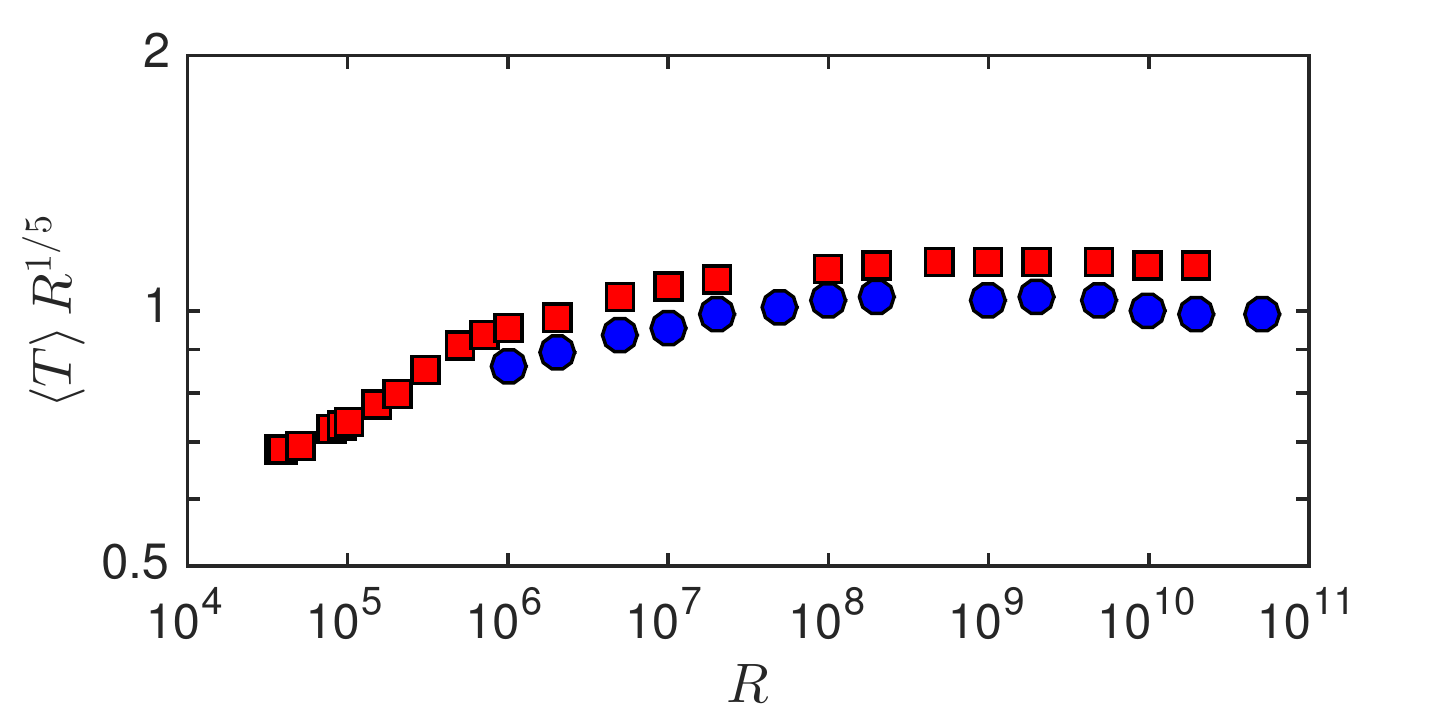}};
\node at (6.4,-2.92) {\includegraphics[width=193pt]{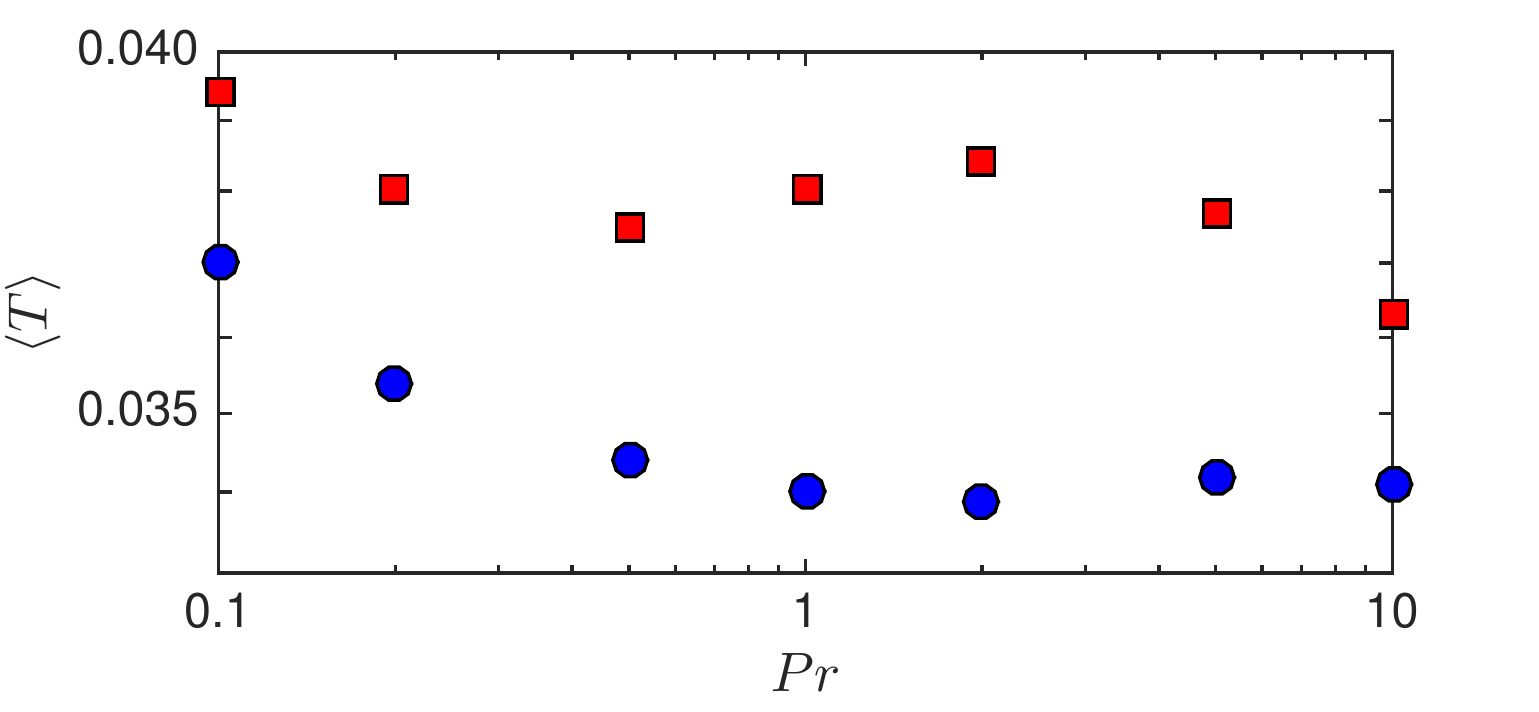}};
\node at (-2.1,-.7) {(a)};
\node at (-2.1,-3.65) {(c)};
\node at (4.35,-.8) {(b)};
\node at (4.35,-3.65) {(d)};
\end{tikzpicture}
\caption{\label{fig:IQs} Variation of mean quantities with $R$ and $Pr$ in 2D (\textcolor[rgb]{1,0,0}{\tiny{$\blacksquare$}}) and 3D (\textcolor[rgb]{0,0,1}{\large{$\bullet$}}) simulations. The fraction of internally produced heat flowing outward across the bottom boundary, $\fB$, is shown (a) for various $R$ with $Pr=1$ and (b) for various $Pr$ with $R=2\e7$. For the same simulations, the dimensionless mean fluid temperature, $\T$, is shown (c) for various $R$ (compensated by $R^{-1/5}$) and (d) for various $Pr$. The data shown are tabulated in the Appendix, except for the 2D data in panels $a$ and $c$, which are from \citet{Goluskin2012}.}
\end{figure}
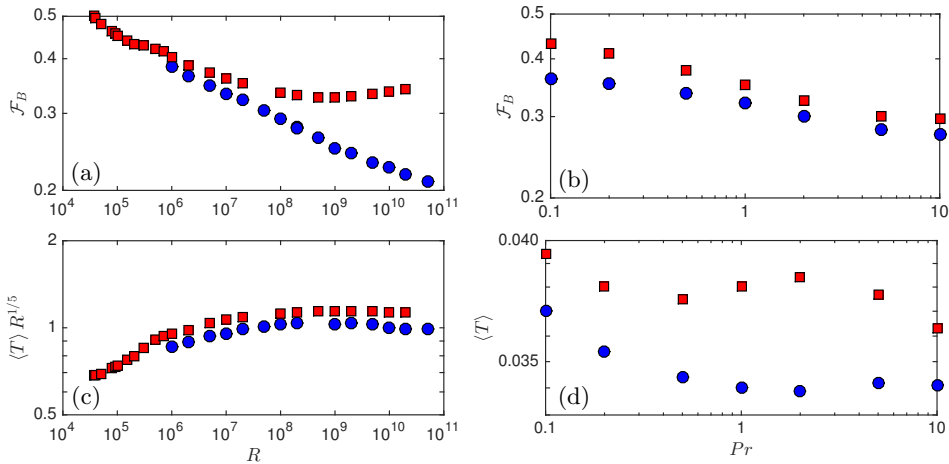 

Figure \ref{fig:IQs}a shows the $R$-dependence of the down-flowing heat fraction, $\fB$, in 2D (\textcolor[rgb]{1,0,0}{\tiny{$\blacksquare$}}) and 3D (\textcolor[rgb]{0,0,1}{\large{$\bullet$}}) simulations with $Pr=1$. As $R$ is raised through moderate values, $\fB$ falls at similar rates in 2D and 3D. At larger $R$, the two cases diverge dramatically. In 2D, $\fB$ stops falling around $R=10^9$ and then slowly rises. In 3D, $\fB$ continues to decay up to the largest $R$ simulated. This decay is steady but quite slow, being approximated well for $R\ge 5\e7$ by $\fB\sim0.80\,R^{0.055}$. It is hard to anticipate whether this decay will persist, let alone to justify its rate. The value of $\fB$ can be viewed as the result of two competing effects: buoyancy-driven mixing of the cold top boundary layer helps heat escape out the top, which lowers $\fB$, while shear-driven mixing of the cold bottom boundary layer helps heat escape out the bottom, which raises $\fB$. Both effects strengthen as $R$ is increased, so the change in $\fB$ depends on which effect strengthens faster.

\begin{figure}
\centering
\begin{tabular}{ccc}
\qquad$R=10^6$ & \qquad$R=10^8$ & \qquad$R=10^{10}$ \\
\includegraphics[height=104pt]{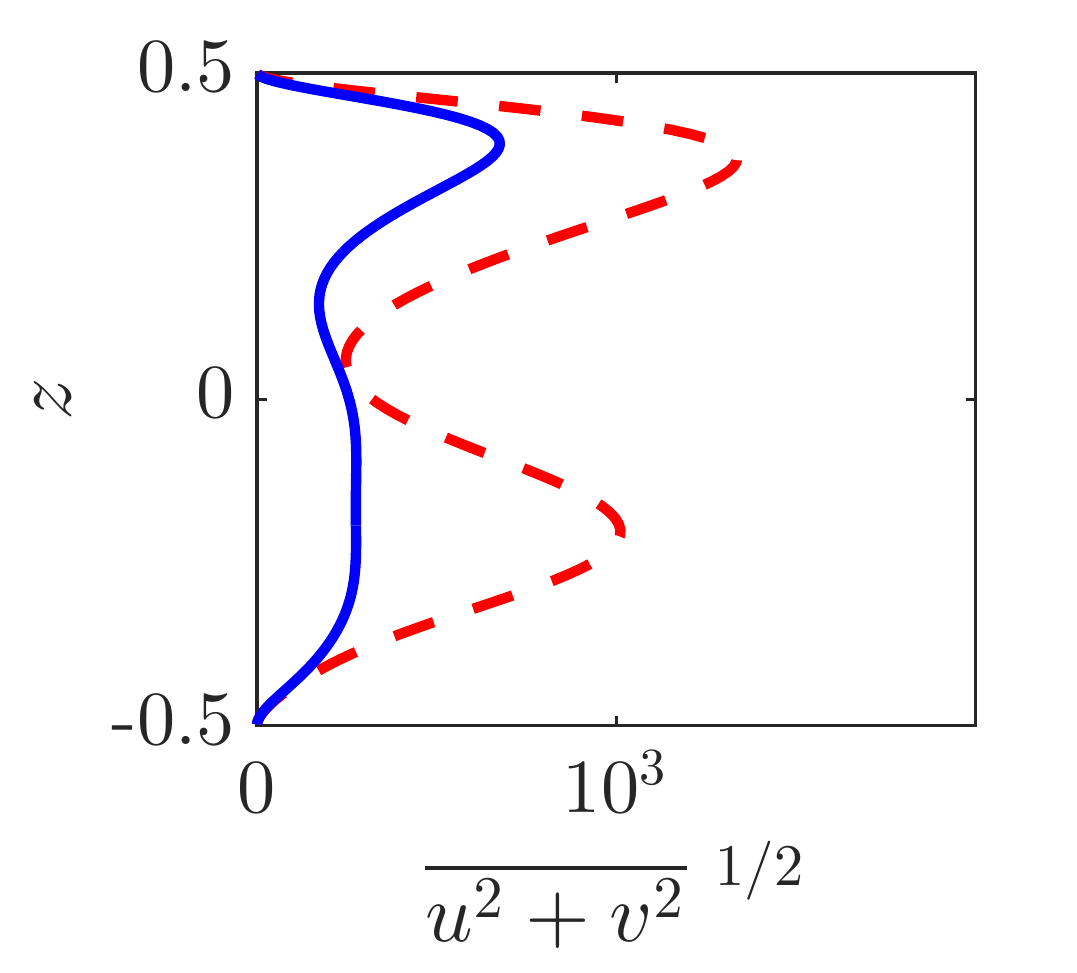} &
\includegraphics[height=104pt]{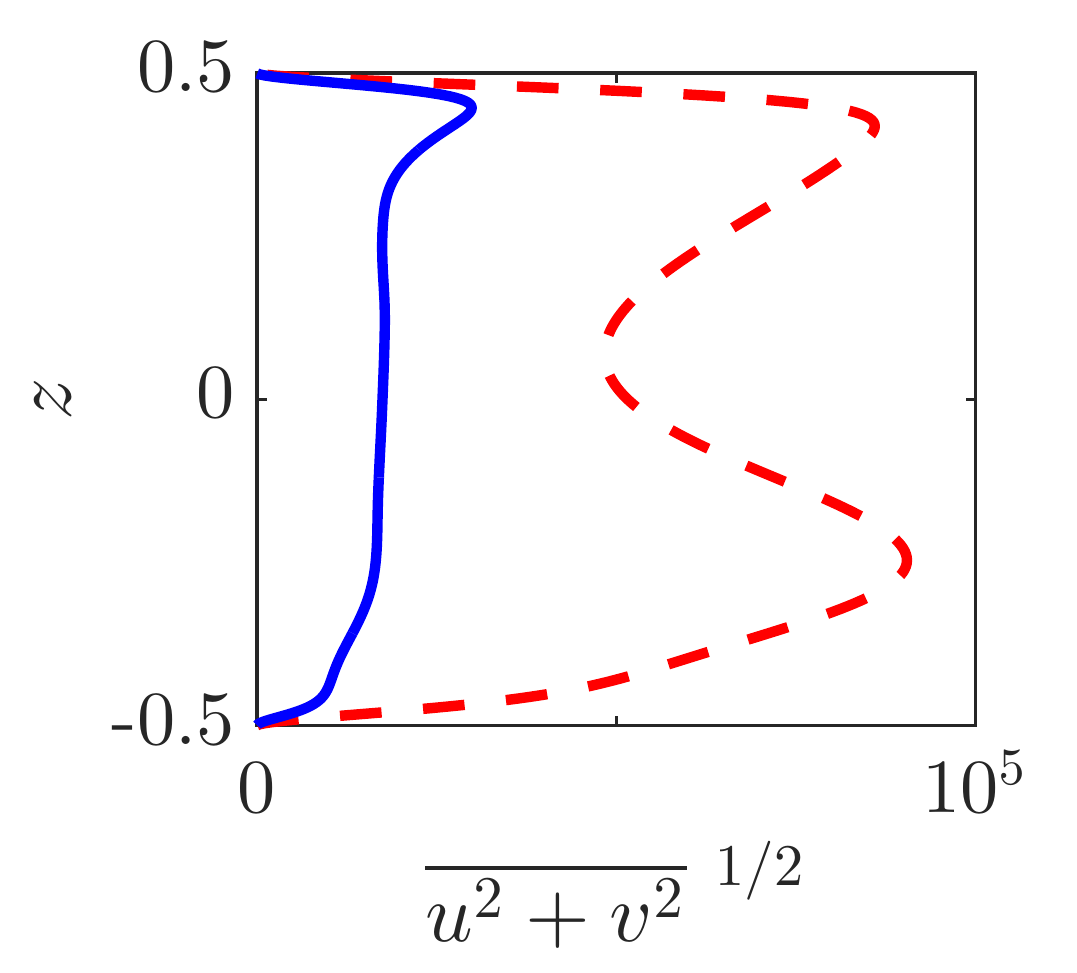} &
\includegraphics[height=104pt]{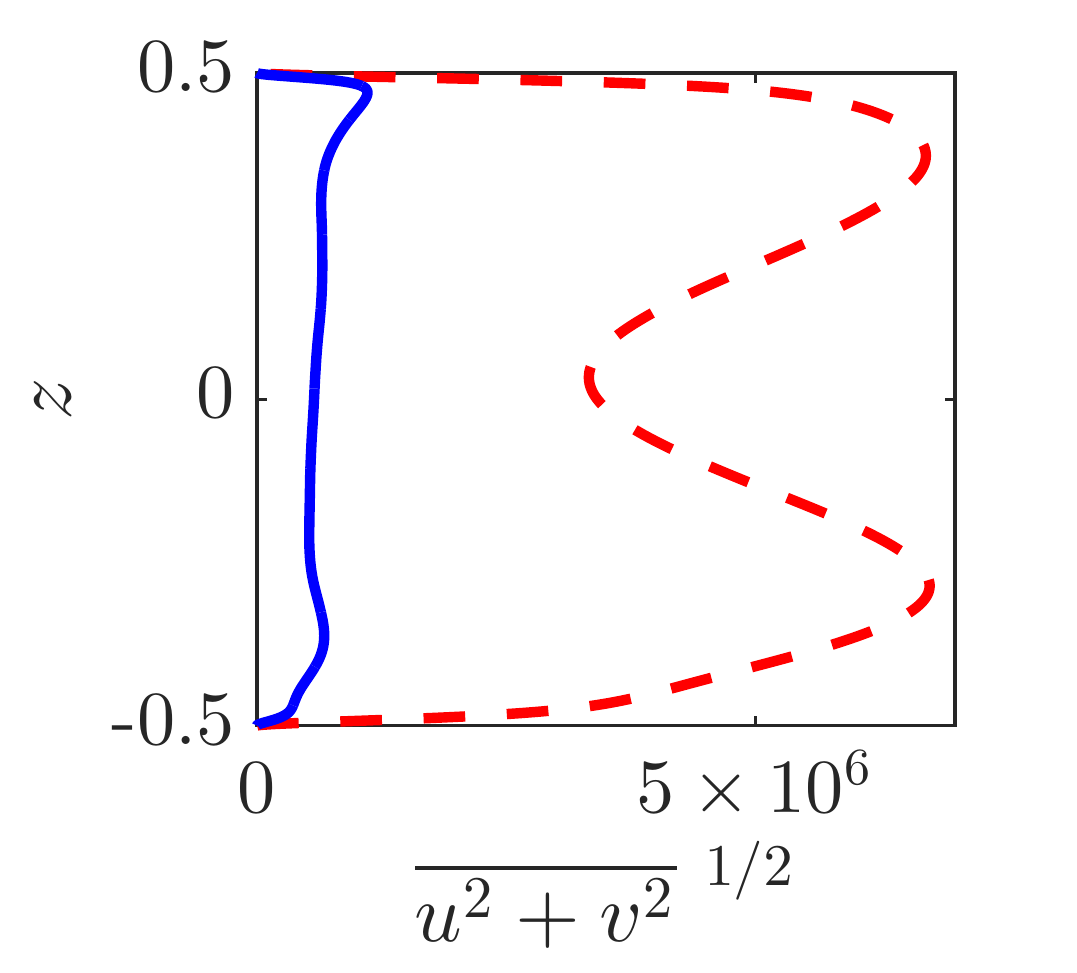} \\
\includegraphics[height=104pt]{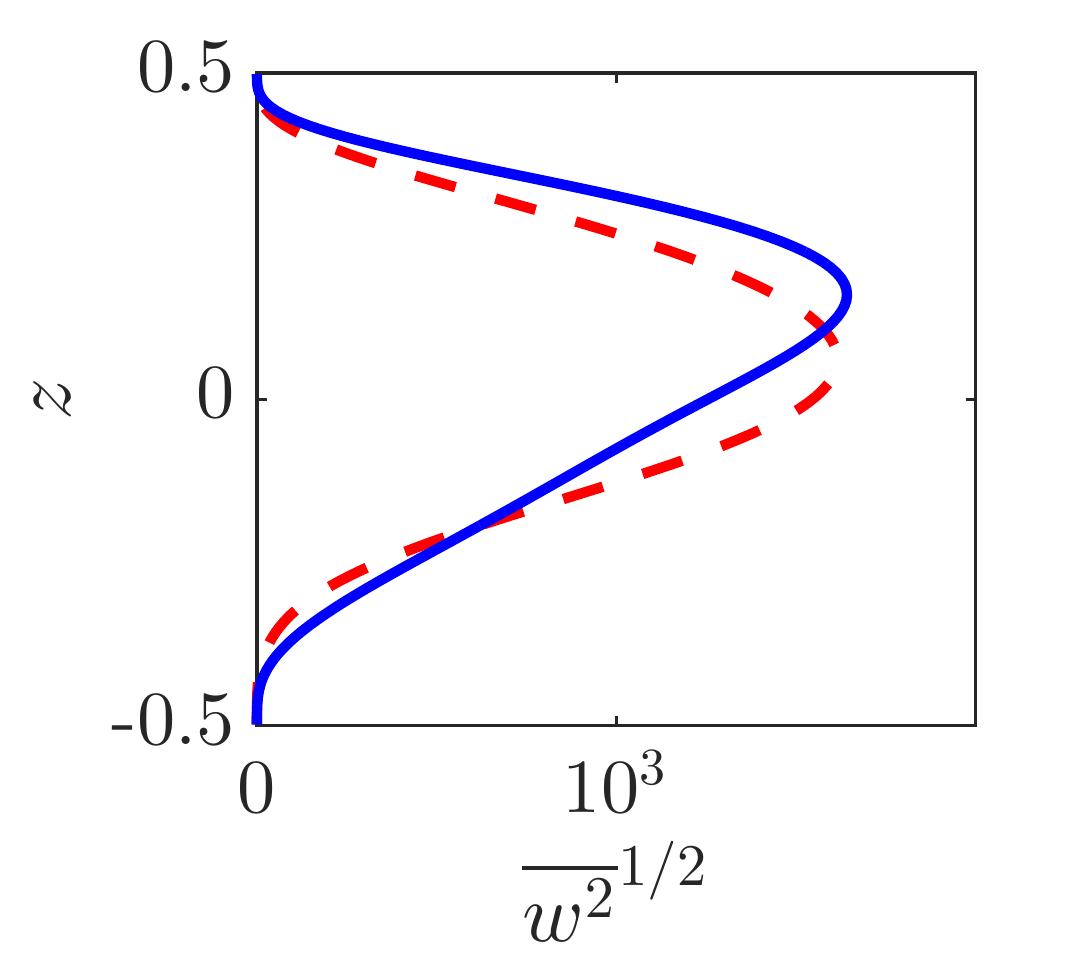} &
\includegraphics[height=104pt]{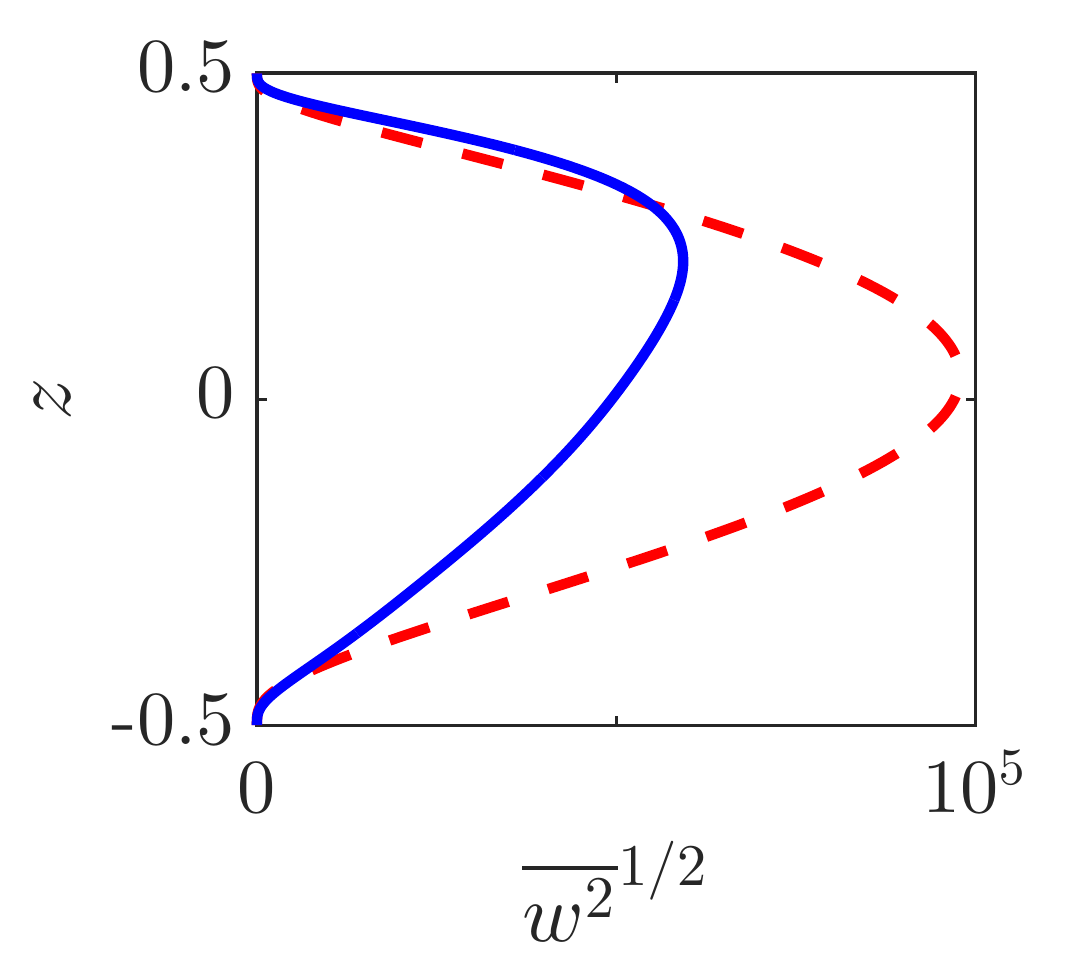} &
\includegraphics[height=104pt]{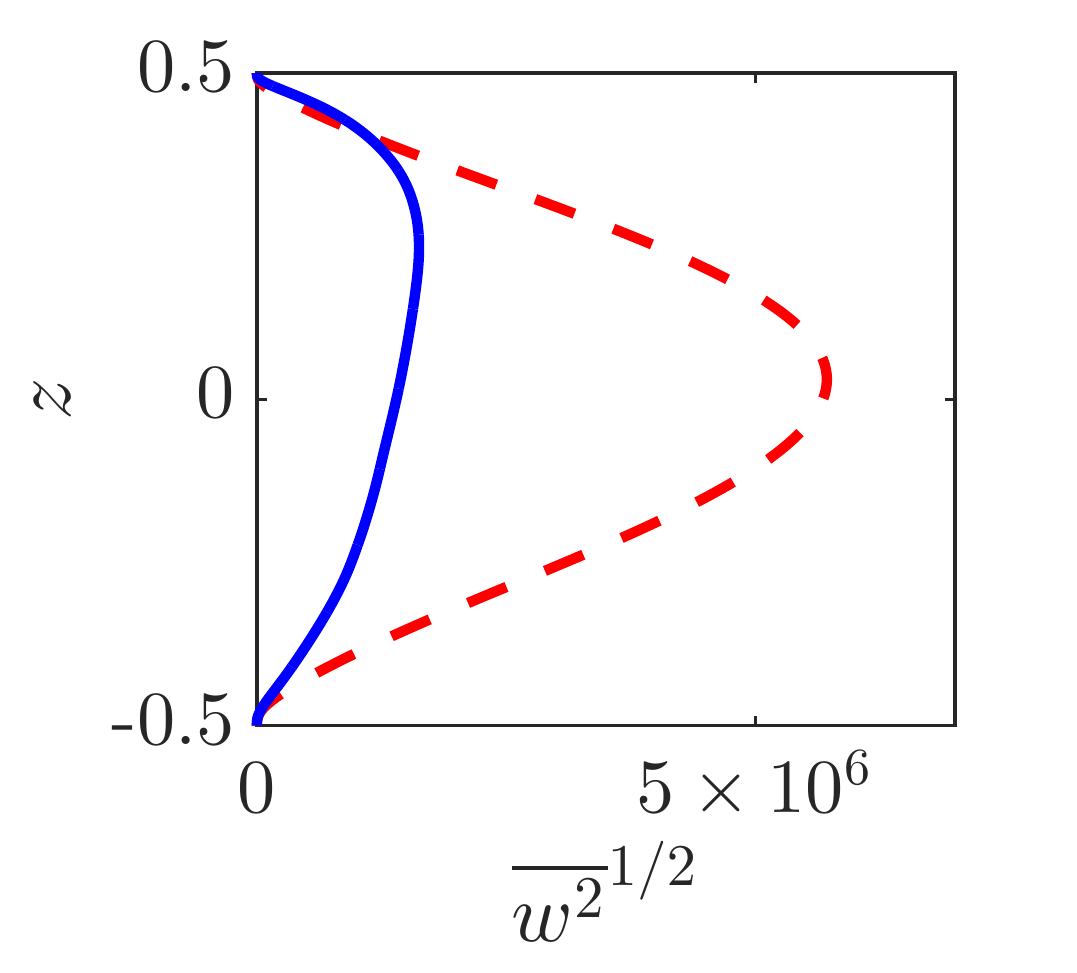}
\end{tabular}
\caption{\label{fig:KE} Mean vertical profiles of root mean square horizontal velocity (top) and vertical velocity (bottom) for 2D ({\color{red}$\dashedrule$}) and 3D ({\color{blue}$\solidrule$}) simulations with $Pr=1$ and $R=10^6,10^8,10^{10}$. In the 2D cases, $v\equiv0$. The 3D profiles are qualitatively consistent with those reported by \citet{Worner1997}.}
\end{figure}

We expect that the emergence of a large-scale circulation (LSC) in our 2D simulations is partly why the bottom boundary layer mixes more effectively in 2D (and thus why $\fB$ is larger in 2D). Such LSC is seen often in RB convection in both 2D and 3D \citep{Ahlers2009}, but mean flows can grow especially strong in 2D, where the vorticity of the plumes is entirely aligned with that of the LSC. Here we have seen no LSC in 3D, although one may arise in larger domains. Figure \ref{fig:KE} shows profiles of root mean square velocities in the horizontal and vertical directions for 2D and 3D simulations at three different $R$. The velocities are stronger in the 2D simulations than in the 3D ones, and this difference increases as $R$ is raised, as in RB convection \citep{VanderPoel2013}. In 3D, the profiles are not only weaker than the 2D ones but are also skewed strongly upward at all $R$, indicating less convective penetration. In 2D, on the other hand, the horizontal velocity profiles become visibly \emph{more} symmetric as convection strengthens, suggesting a strong LSC that penetrates to the bottom boundary layer. The greater asymmetry of velocity profiles in 3D is consistent with the greater asymmetry of heat fluxes in 3D that is reflected by smaller values of $\fB$ in figures \ref{fig:IQs}(a-b).

Figure \ref{fig:IQs}c shows the dependence of $\T$ on $R$, compensated by $R^{-1/5}$. When $R$ is large, $\T$ decays like $R^{-1/5}$ in both 2D and 3D, as reflected by nearly flat trends in the compensated plot. (Fits to data for $R\ge10^8$ give $\T\sim1.13\,R^{-0.20}$ in 2D and $\T\sim1.11\,R^{-0.20}$ in 3D. The 3D fit, while less good than the 2D fit, is robust; excluding any two data points gives an exponent of either $-0.20$ or $-0.21$.) The decay exponent of $-1/5$ corresponds to the \emph{dimensional} mean temperature growing with the heating rate like $H^{4/5}$.

The maximum of $\oT(z)$ on the interior, $\Tm$, is necessarily larger than $\T$, but the two values grow closer as $R$ is raised. For instance, in the 3D simulations $\Tm$ is 28\% larger than $\T$ when $R=10^6$ but only 8\% larger when $R=10^{10}$ (cf.\ table \ref{table2} in the Appendix). This reflects the flattening of temperature profiles that is evident in figure \ref{fig2}. The decay of $\Tm$ is thus slightly faster than that of $\T$, being fit well by $\Tm\sim1.62\,R^{-0.22}$ in 3D. This exponent is consistent with past experiments \citep{Kulacki1972, Jahn1974, Mayinger1975, Ralph1977, Lee2007} and 3D simulations \citep{Worner1997}, where $\Tm$ falls at rates between $R^{-0.18}$ and $R^{-0.22}$ \citep{Goluskin2015a}. At very large $R$, we expect $\Tm\approx\T$ once $\oT(z)$ becomes very flat outside of negligibly thin boundary layers.

Figures \ref{fig:IQs}b and \ref{fig:IQs}d show the $Pr$-dependence of $\fB$ and $\T$, respectively, with $R=2\e7$. When $\Pr\to0$, deviations from the static temperature profile cannot be maintained, so we expect $\fB$ and $\T$ to approach their static values in both 2D and 3D, much like the Nusselt number in RB convection. When $Pr\to\infty$, we expect both quantities to saturate at intermediate values as solutions of the Boussinesq equations approach solutions of the infinite-$Pr$ Boussinesq equations \citep{Wang2004, Wang2008}. The findings of \citet{Schmalzl2004} on RB convection suggest that the infinite-$Pr$ limits of mean quantities will be similar in 2D and 3D. Behaviour is more complicated at intermediate $Pr$, where the deviation between 2D and 3D flows is expected to be largest. Although $\fB$ varies monotonically over the $Pr$ range simulated, $\T$ does not. Below we relate this non-monotonicity of $\T$ to that of the RB Nusselt number.

Quantities like $\fB$ that measure vertical asymmetry seem to have no analogues in RB convection. On the other hand, the dimensionless ratio $1/\T$, which is proportional to the ratio of the heating rate $H$ to the \emph{dimensional} mean temperature, behaves much like the Nusselt number in RB convection. This motivates us to define a Nusselt number, $N$, for IH convection and consider its dependence on a diagnostic Rayleigh number, $Ra$, that differs from the control parameter~$R$.

The quantities $N$ and $Ra$ should be defined to convey the strengths of convective transport and thermal forcing, respectively. When comparing RB models with different thermal boundary conditions, it is most useful to define $N$ as the ratio of total vertical heat flux to conductive vertical heat flux in the flow, and to define $Ra$ like $R$ but using a temperature scale that is characteristic of the developed flow \citep{Otero2002, Johnston2009, Wittenberg2010}. In the present model, we cannot normalize $N$ using the mean vertical conduction, which is zero, but we can instead consider \emph{outward} heat flux: upward flux above the height where $\Tm$ occurs, plus downward flux below it \citep{Goluskin2012}. Total outward flux is fixed on average, but conductive outward flux is proportional to $\Tm$, so $N$ should be proportional to $1/\Tm$. The dimensional version of $\Tm$ can be used also as the temperature scale defining $Ra$, yielding
\begin{align}
N &:= \frac{1}{8\,\Tm}, & Ra &:= \frac{R}{N}, \label{eq: N}
\end{align}
where $N=1$ and $Ra=R$ in the static state. Alternatively, choosing $\T$ as the temperature scale instead of $\Tm$ gives the similar definitions
\begin{align}
\wt N &:= \frac{1}{12\T}, & \wt{Ra} &:= \frac{R}{\wt N}. \label{eq: Nwt}
\end{align}
Further discussion of how to define Nusselt-number-like quantities in IH convection is given by \citet{Goluskin2015a}.

\begin{figure}
\centering
\begin{tikzpicture}
\node at (0,0) {\includegraphics[width=192pt]{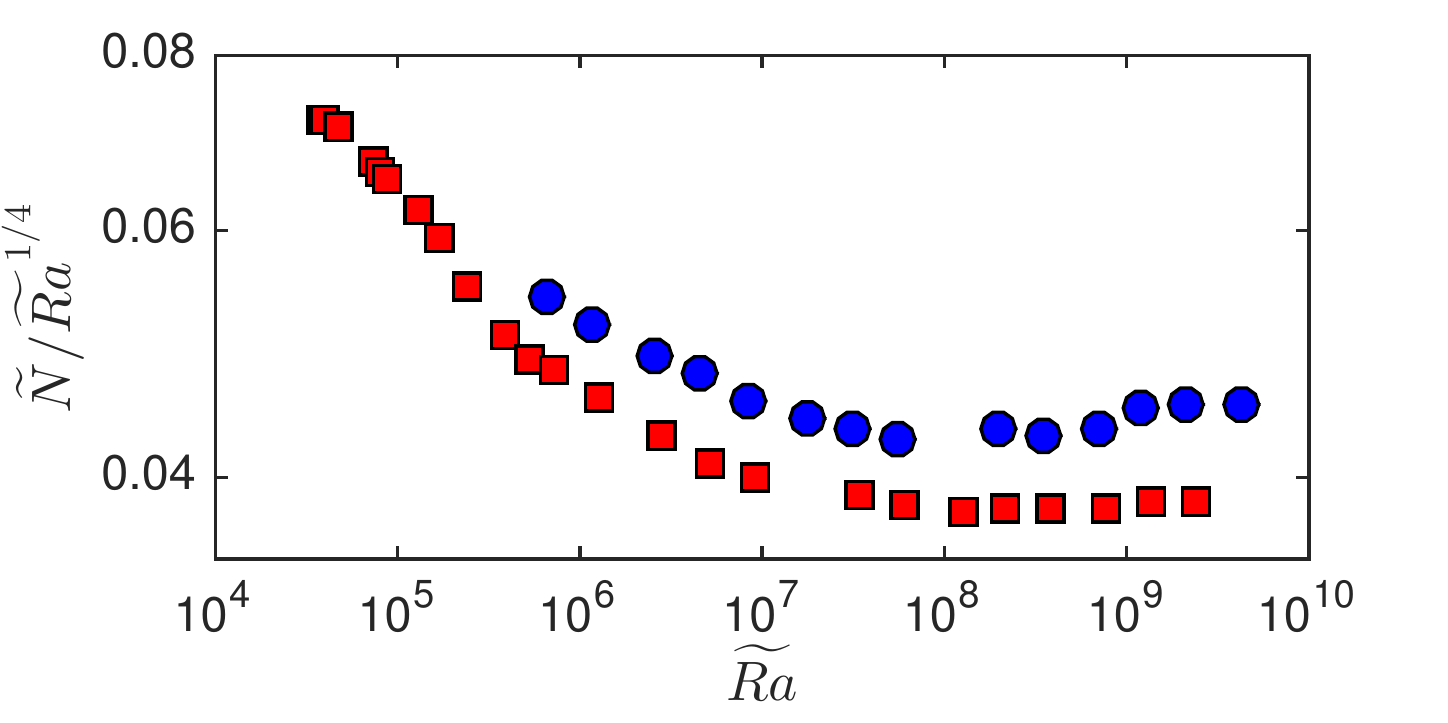}};
\node at (6.5,0) {\includegraphics[width=189pt]{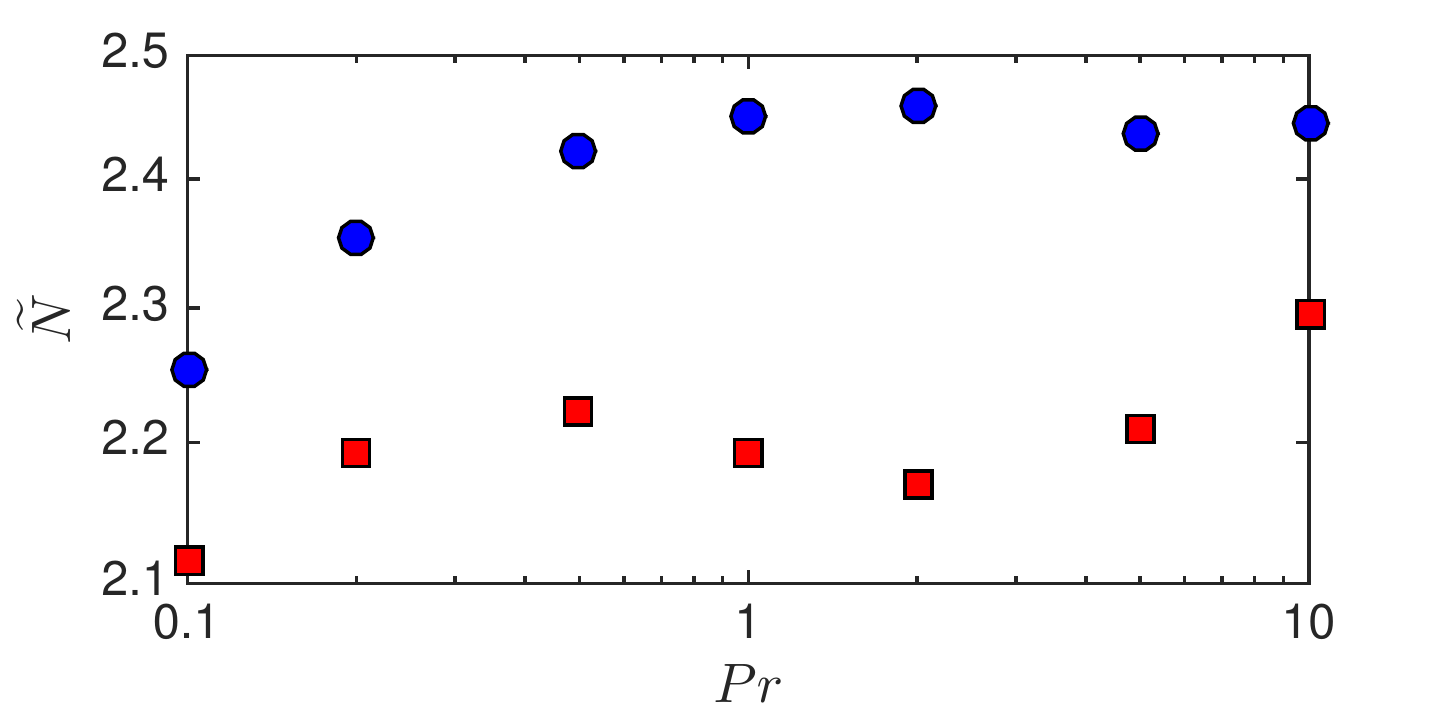}};
\node at (-2.1,-.7) {(a)};
\node at (4.4,-.8) {(b)};
\end{tikzpicture}
\caption{\label{fig:N} Variation of $\wt{N}$ (a) with $\wt{Ra}$ (compensated by $\wt{Ra}^{1/4}$) and (b) with $Pr$ in 2D (\textcolor[rgb]{1,0,0}{\tiny{$\blacksquare$}}) and 3D (\textcolor[rgb]{0,0,1}{\large{$\bullet$}}) simulations. The $\wt{N}$ and $\wt{Ra}$ values are calculated according to \eqref{eq: Nwt} from the $\T$ values represented in figures \ref{fig:IQs}c and \ref{fig:IQs}d.}
\end{figure} 

Nusselt and Rayleigh numbers defined according to \eqref{eq: N} or \eqref{eq: Nwt} display a parameter-dependence similar to the Nusselt number in RB convection. For instance, the fits to $\Tm$ and $\T$ reported above for 3D simulations with $Pr=1$, when formulated in terms of Nusselt numbers, become $N\sim0.038\,Ra^{0.28}$ and $\wt N\sim0.039\,\wt{Ra}^{0.26}$, respectively. These exponents are within the ranges seen in RB experiments at similar parameters, although they are slightly smaller than usual \citep{Grossmann2000, Ahlers2009}. The approximate scaling of $\T$ like $R^{-1/5}$ corresponds to $\wt N$ scaling like $\wt{Ra}^{1/4}$. The closeness of the data to this scaling is shown by figure \ref{fig:N}a, which re-expresses the $\T$ data of figure \ref{fig:IQs}c in terms of (compensated) $\wt{N}$ and $\wt{Ra}$. Moreover, scaling arguments like those of \citet{Malkus1954}, \citet{Kraichnan1962}, \citet{Spiegel1971b}, or \citet{Grossmann2000} predict the same Nusselt number scalings for IH convection as for RB convection, provided that Nusselt and Rayleigh numbers are defined by \eqref{eq: N} or \eqref{eq: Nwt}. Even mathematical bounds share the same scalings; the bounds (\ref{eq: T bounds}) on $\T$ correspond to $1\le\wt N<0.025\wt{Ra}^{1/2}$, and upper bounds with this same exponent have been proven for 3D RB convection with various boundary conditions \citep{Constantin1996, Plasting2003, Otero2002, Wittenberg2010}. Finally, the parallels with the RB Nusselt number extend also to the $Pr$-dependence of $\wt N$, shown in figure \ref{fig:N}b. Both the 2D and 3D trends in this figure resemble the analogous results for RB convection \citep[cf.\ figure 6b of][]{VanderPoel2013}.

\section{Conclusions}
\label{sec: con}

We have explored the influence of Rayleigh number, Prandtl number, and spatial dimension on penetrative IH convection. In dimensional terms, the mean fluid temperature grows with the heating rate ($H$) proportionally to $H^{4/5}$. The ratio of $H$ to this temperature is found to behave much like the Nusselt number in RB convection, and we have proposed an analogy in which the $H^{4/5}$ scaling corresponds to the Nusselt number growing like the 1/4 power of the Rayleigh number. It remains to be seen whether this analogy will survive a wider exploration of parameter space. In particular, the crossovers between different scalings seen in the RB system \citep{Stevens2013} are yet to be found in IH convection. We have examined the fraction of heat flowing outward across the bottom boundary, $\fB$, that initially falls as convection strengthens, and which is greatly influenced by the extent to which convection penetrates the stable bottom boundary layer. At the largest $R$ simulated, $\fB$ continues to fall toward zero in 3D but has reversed its fall in 2D. This disparity warns against using 2D simulations to approximate 3D convection when it is penetrative. Although our simulations provide a clearer picture of how $\fB$ behaves in moderately strong convection, this behaviour is still not well understood. No simple arguments have been found to predict the parameter-dependence of $\fB$, and even the fate of this fraction in the infinite-Rayleigh-number limit remains obscured. The ability to answer such a seemingly simple question is surely a prerequisite to understanding internally heated convection in nuclear reactors, planets, and stars.

\section*{Acknowledgements}
We thank D.\ Lohse, R.\ Verzicco, R.\ Ostilla M\'onico, \mbox{C.\ R.}\ Doering, O.\ Zikanov, \mbox{E.\ A.}\ Spiegel, and the anonymous referees for their helpful remarks, as well as H. Johnston for initiating our collaboration. During much of this work, D.G.\ was supported by NSF award PHY-1205219, and E.P.\ was supported by FOM and NWO grant SH-202.

\appendix

\section{Data and convergence studies}

\begin{table}
\begin{center}
\caption{\label{table1} Details of the 3D finite difference simulations represented in figures \ref{fig2}-\ref{fig:N}. The columns from left to right indicate the control parameters ($R$, $Pr$, $\Gamma$), the spatial resolution ($N_x \times N_y \times N_z$), the number of grid points in the steeper (top) thermal boundary layer (whose thickness is approximated as $\T /[1/2 + \wT]$), the ratio of the average Kolmogorov length scale ($\eta= Pr^{1/2}/[R\wT]^{1/4}$) to the maximum grid length ($\max\{\delta_x,\delta_z\}$), the post-transient simulation time in thermal units ($\tau=d^2/\kappa$), and the directly computed averages $\wT$, $\T$, and $\Tm$. The quantity $\wT$ gives the fraction $\fB$ since $\fB=1/2-\wT$.}
    \begin{tabular}{c c c c c c c c c c c }
    R & Pr & $\Gamma$ &$N_x \times N_y \times N_z$ &  $N_{BL}$ & $\frac{\eta}{\max(\delta_x,\delta_z)}$ &  $\tau$ & $\wT$ & $\T$ & $\Tm$ \\[2pt]
    $10^6$            			& 1 	& $\pi$ & $288 \times 288 \times 144$   & 19 & 6.0 & 1.1 & 0.117 & 0.0547 & 0.0698\\
    $2\e6$            	& 1 	& $\pi$ & $288 \times 288 \times 144$   & 17 & 5.2 & 0.7 & 0.135 & 0.0490 & 0.0616 \\
    $5\e6$            	& 1 	& 2 	& $240 \times 240 \times 144$   & 15 & 5.6 & 1.0 & 0.153 & 0.0427 & 0.0522 \\
    $10^7$            			& 1 	& 2 	& $240 \times 240 \times 144$   & 12 & 4.8 & 0.5 & 0.168 & 0.0380 & 0.0457 \\
    $2\e7$            	& 1 	& 2 	& $240 \times 240 \times 144$   & 11 & 4.2 & 0.5 & 0.179 & 0.0338 & 0.0400 \\
    $5\e7$            	& 1 	& 1.5 	& $288 \times 288 \times 216$   & 18 & 5.5 & 0.9 & 0.195 & 0.0291 & 0.0336 \\
    $10^8$            	& 1 	& 1.5 	& $360 \times 360 \times 240$   & 18 & 6.0 & 0.3 & 0.208 & 0.0258 & 0.0294\\
    $2\e8$             & 1 	& 1.5 	& $360 \times 360 \times 240$   & 16 & 5.2 & 0.4 & 0.222 & 0.0225 & 0.0253 \\
    $5\e8$             & 1 	& 1.5 	& $432 \times 432 \times 288$   & 18 & 5.2 & 0.5 & 0.238 & 0.0189 & 0.0209 \\
    $10^9$            & 1 	& 1 	& $480 \times 480 \times 360$   & 22 & 7.5 & 0.3 & 0.252 & 0.0163 & 0.0180\\
    $2\e9$             & 1 	& 1 	& $480 \times 480 \times 384$   & 21 & 6.2 & 0.2 & 0.257 & 0.0142 & 0.0154 \\
    $5\e9$             & 1 	& 1 	& $432 \times 432 \times 432$   & 21 & 4.9 & 0.1 & 0.267 & 0.0117 & 0.0126 \\
    $10^{10}$        & 1 	& 1 	& $576 \times 576 \times 576$   & 27 & 5.8 & 0.1 & 0.274 & 0.0100 & 0.0108 \\
    $2\e{10}$          & 1 	& 1 	& $720 \times 720 \times 720$   & 35 & 6.3 & 0.1 & 0.283 & 0.0087 & 0.0092 \\
    $5\e{10}$          & 1 	& 1 	& $864 \times 864 \times 864$   & 37 & 6.3 & 0.1 & 0.315 & 0.0070 & --- \\
    $2\e7$          	& 0.1 	& 2 	& $360 \times 360 \times 240$   & 26 & 1.9 & 4.1 & 0.138 & 0.0370 & 0.0448\\
    $2\e7$          	& 0.2 	& 2 	& $360 \times 360 \times 240$   & 25 & 2.8 & 3.8 & 0.147 & 0.0355 & 0.0424 \\
    $2\e7$          	& 0.5 	& 2 	& $360 \times 360 \times 240$   & 24 & 4.4 & 2.3 & 0.163 & 0.0344 & 0.0408 \\
    $2\e7$          	& 2 	& 2 	& $360 \times 360 \times 240$   & 23 & 8.9 & 2.1 & 0.199 & 0.0339 & 0.0400 \\
    $2\e7$          	& 5 	& 2 	& $360 \times 360 \times 240$   & 23 & 14  & 2.0 & 0.218 & 0.0342 & 0.0403 \\
    $2\e7$          	& 10 	& 2 	& $360 \times 360 \times 240$   & 22 & 20  & 5.1 & 0.225 & 0.0341 & 0.0402
  \end{tabular}
\end{center}
\end{table}

Table \ref{table1} provides details on the 3D simulations whose results appear in figures \ref{fig2}--\ref{fig:N}. These details include time spans, meshes, boundary layer resolutions, and Kolmogorov length scales. Time averages in the tabulated values of $\wT$ and $\T$ are converged to the precision shown, or nearly so. In all cases with $Pr=1$ and $R\le2\e8$, simulations have been repeated using the {\tt nek5000} code, and the resulting values of $\wT$ and $\T$ agree with the tabulated values to within 1\%. Table \ref{table2} gives details on the 2D simulations where $Pr$ was varied with $R=2\times10^7$.

\begin{table}
\begin{center}
\caption{\label{table2} Details of the 2D finite difference simulations at various $Pr$ represented in figures \ref{fig:IQs}b and \ref{fig:IQs}d. In all cases, $R=2\times10^7$, $\Gamma=12$, and the spatial resolution is $3072\times256$. The quantity $\wT$ gives the fraction $\fB$ since $\fB=1/2-\wT$.}
    \begin{tabular}{c c c c c c c c }
    Pr & $\wT$ & $\T$ \\[2pt]
    0.1 	& 0.0677 & 0.0394 \\
    0.2 	& 0.0891 & 0.0380 \\
    0.5 	& 0.122~ & 0.0375 \\
    1 	& 0.148~ & 0.0380 \\
    2 	& 0.175~ & 0.0384 \\
    5 	& 0.200~ & 0.0377 \\
    10 	& 0.204~ & 0.0363
  \end{tabular}
\end{center}
\end{table}	

\begin{table}
\centering
\caption{\label{table3} Convergence of $\T$ and $\wT$ in 3D simulations as the horizontal extent in both directions, $\Gamma$, is raised with $Pr=1$ and various fixed $R$. Computations were carried out using {\tt nek5000} for the two lower $R$ values and using the finite difference code for the higher $R$ values.}
\begin{tabular}{cllllllllllll}
R & $5\e5$ &&& $2\e6$ &&& $2\e7$ && $10^9$ \\
$\Gamma$ & 2 & $\pi$ & 4.5 & 1.5 & 2 & $\pi$ & 1.5 & 2 & 1 & 1.5 \\
$\wT$ & 0.0939 & 0.1000 & 0.0997 & 0.126 & 0.134 & 0.134 & 0.179 & 0.177 & 
	0.247 & 0.247 \\
$\T$ & 0.0626 & 0.0604 & 0.0606 & 0.0505 & 0.0492 & 0.0491 & 0.0340 & 0.0343 &
	0.0164 & 0.0164
\end{tabular}
\end{table}

Table \ref{table3} illustrates the convergence of $\T$ and $\wT$ as the aspect ratio $\Gamma$ is increased with $Pr=1$. Convergence evidently occurs at smaller $\Gamma$ when $R$ is larger. Since we have varied $\Gamma$ systematically only with $Pr=1$, further work is needed to explore the effects of $\Gamma$ at other $Pr$. The RB simulations of \citet{Hartlep2005} suggest that larger $Pr$ might require larger $\Gamma$ for $\T$ and $\wT$ to converge.

\bibliographystyle{jfm}
\bibliography{Penetrative_IH_convection.bbl}

\end{document}